\documentclass[preprint]{JHEP3}

\def\bea{\begin{eqnarray}}
\def\eea{\end{eqnarray}}
\def\bec{\begin{center}}
\def\ec{\end{center}}

\def\beq{\begin{equation}}
\def\eeq{\end{equation}}

\def\f{\frac}

\def\f#1#2{\frac{#1}{#2}}

\def\bea{\begin{eqnarray}}
\def\eea{\end{eqnarray}}
\def\beq{\begin{equation}}
\def\eeq{\end{equation}}
\preprint{KAIST-TH 2003/05, KUNS-1841}
\title{Supersymmetry Breaking in Warped Geometry}
\author{Kiwoon Choi\\Department of Physics,
Korea Advanced Institute of Science and Technology, \\
Daejeon 305-701, Korea 
\\ E-mail: \email{kchoi@hep.kaist.ac.kr}}
\author{Do Young Kim\\Department of Physics,
Korea Advanced Institute of Science and Technology, \\
Daejeon 305-701, Korea 
\\ E-mail: \email{kimdoyoung@hep.kaist.ac.kr}}
\author{Ian-Woo Kim\\Department of Physics,
Korea Advanced Institute of Science and Technology, \\
Daejeon 305-701, Korea 
\\ E-mail: \email{iwkim@hep.kaist.ac.kr}}
\author{Tatsuo Kobayashi\\Department of Physics, Kyoto University,
Kyoto 606-8502, Japan
\\ E-mail: \email{kobayash@gauge.scphys.kyoto-u.ac.jp}}

\abstract{
We examine the soft supersymmetry breaking parameters in  
supersymmetric theories on a slice of ${\rm AdS}_5$ 
which generate the hierarchical Yukawa couplings by dynamically
localizing the bulk matter fields in extra dimension.
Such models can be regarded as the AdS dual of the recently studied 
4-dimensional models which contain a supersymmetric CFT
to generate the hierarchical Yukawa couplings.
In such models, if  supersymmetry breaking is
mediated by the bulk radion superfield
and/or some brane chiral superfields,
potentially dangerous flavor-violating soft parameters
can be naturally suppressed,
thereby avoiding 
the SUSY flavor problem.
We present some models of radion-dominated supersymmetry
breaking which yield a highly predictive form of soft parameters
in this framework.
}

\begin{document}
\section{Introduction}

Warped extra dimension can provide a rationale 
for various hierarchical structures in
particle physics. For instance, in 5-dimensional (5D) theory on a
slice of $\mbox{AdS}_5$ with AdS curvature $k$ and orbifold radius
$R$, the small warp factor $e^{-\pi kR}$  can generate the huge
hierarchy between the 4D Planck scale ($M_{Pl}\sim 10^{18}$ GeV)
and the weak scale ($M_W\sim 10^2$ GeV) \cite{Randall:1999ee},
and/or the hierarchical quark and lepton masses
\cite{Gherghetta:2000qt,Huber:2000ie}, and/or small neutrino
masses \cite{Grossman:1999ra}. 
The underlying dynamics for small $M_W/M_{Pl}$
and small fermion masses 
was the localization of
gravity and fermion zero modes in extra dimension.

In this paper, we wish to examine
soft supersymmetry (SUSY) breaking parameters 
in  models with warped extra dimension
in which the hierarchical Yukawa couplings are generated 
by localizing the bulk matter fields in extra dimension
\cite{previous}.
To this end, we consider supersymmetric 5D theory on a slice of 
$\mbox{AdS}_5$
in which all quark and lepton fields arise from 5D bulk 
hypermultiplets \cite{Gherghetta:2000qt}. 
The graviphoton of the model can be gauged appropriately
to have nonzero bulk cosmological constant
and hypermultiplet masses which would
make the gravity zero mode localized at 
one boundary ($y=0$) and  the matter zero modes  
localized at any of the two boundaries
($y=0$ or $\pi$).
In this type of models,
the small warp factor and
the small Yukawa couplings have a common origin,
i.e. the dynamical localization of zero modes associated
with the gauging of graviphoton. As a result, 
the small Yukawa couplings are typically given by
$y\sim e^{-c\pi kR}$ where $c$ is a constant of order unity,
which would require that
the radion is stabilized at a value
giving the warp factor: $e^{-\pi kR}\approx 10^{-2}-10^{-5}$
and the corresponding 
Kaluza-Klein (KK) scale:
$M_{KK}\approx k e^{-\pi kR}\sim 10^{16}-10^{13}$
GeV.
This type of models can be regarded as the AdS dual of the 
recently studied 4D models \cite{Nelson:2000sn} which contain
a supersymmetric CFT sector generating the 
Yukawa hierarchy through  renormalization group
evolution.

It has been noted that supersymmetric CFT dynamics can provide
also a mechanism to suppress dangerous flavor-violating soft parameters
\cite{Nelson:2000sn}.
This suggests that a similar suppression
of flavor-violation can take place in $\mbox{AdS}_5$ models also.
As we will see,  flavor-violating soft parameters
in $\mbox{AdS}_5$ models are indeed suppressed under a 
reasonable assumption
on SUSY breaking.
Furthermore, if SUSY is broken by 
the $F$-term of the radion superfield or equivalently
by the Scherk-Schwarz mechanism \cite{Marti:2001iw,Kaplan:2001cg},
the model leads to a concrete prediction
on soft parameters which can be tested
by future experiments.
In this paper, we first discuss general forms of 
soft parameters in supersymmetric $\mbox{AdS}_5$ model,
and later present some models of radion-dominated
SUSY breaking in $\mbox{AdS}_5$
which can pass all constraints on flavor-violation
without severe fine tuning of parameters.
Most of our results on the Yukawa couplings and
soft parameters  can be applied
also to the models 
with dynamically localized bulk matter fields
in {\it flat} extra dimension.
For this, one can simply take a limit
that the AdS curvature becomes zero, 
while the hypermultiplet masses remain to be nonzero.

\medskip

\section{Yukawa couplings and
Soft parameters in Supersymmetric $\mbox{AdS}_5$ model}

To proceed, let us consider a generic 5D gauge theory
coupled to the minimal 5D supergravity (SUGRA) on
$S^1/Z_2$. The action of the model is given by
\cite{Gherghetta:2000qt,Ceresole:2000jd,Altendorfer:2000rr} \bea \label{5daction1}
S = \int d^5x&&\sqrt{-G} \,\left[\frac{1}{2}\left(\, {\cal
R}+\bar{\Psi}^i_M\gamma^{MNP}D_N\Psi_{iP}-\frac{3}{2}C_{MN}C^{MN}
+12k^2\,\right) \right.\nonumber \\
&&+\frac{1}{{g}_{5a}^2}
\left(-\frac{1}{4}F^{aMN}F^a_{MN}
+\frac{1}{2}D_M\phi^aD^M\phi^a
+\frac{i}{2}\bar{\lambda}^{ai}\gamma^MD_M\lambda^a_i
\right)\nonumber \\
&&\left.+|D_Mh_I^i|^2+i\bar{\Psi}_I\gamma^MD_M\Psi_I+
ic_Ik\epsilon(y)\bar{\Psi}_I\Psi_I +... \right]
\eea
where ${\cal R}$ is the 5D Ricci scalar, $\Psi^i_M$ ($i=1,2$) are
the symplectic Majorana gravitinos,
$C_{MN}=\partial_MB_N-\partial_NB_M$ is the graviphoton field strength, and
$y$ is the 5th coordinate with a fundamental range
$0\leq y\leq\pi$.
Here $\phi^a,A_M^a$ and $\lambda^{ia}$ are 5D scalar, vector and
symplectic Majorana
spinors constituting a 5D vector multiplet,
$h_I^i$ and $\Psi_I$  are 5D scalar and Dirac spinor constituting
the $I$-th hypermultiplet
with kink mass $c_Ik\epsilon(y)$.
The AdS curvature $k$ and the kink mass 
$c_Ik$ are related with the gauging of graviphoton
as indicated by the following covariant derivatives \cite{Ceresole:2000jd,Choi:2002wx}:
\bea
\label{gauging}
&& D_M h^i_I=\partial_M h^i_I-i\left(
\frac{3}{2}(\sigma_3)^i_j-c_I\delta^i_j\right)k\epsilon(y)B_M h^j_I+...,
\nonumber \\
&& D_M\Psi_I=\partial_M\Psi_I+ic_Ik\epsilon(y)B_M\Psi_I+...,
\nonumber \\
&& D_M\lambda^{ai}=\partial_M\lambda^{ai}-i\frac{3}{2}
(\sigma_3)^i_j k\epsilon(y)B_M\lambda^{aj}+...,
\eea
where the ellipses stand for other gauge couplings.
Note that we set the 5D Planck mass
$M_5=1$ and all dimensionful parameters, e.g.
the 5D gauge coupling $g_{5a}$ and the AdS curvature
$k$, are defined in this unit.
Although not required within 5D SUGRA, it is not
unreasonable to assume that the graviphoton
charges ($c_I$) are {\it quantized} in an appropriate
unit, i.e. $c_I/c_J$ are rational numbers, which we will
adopt throughout this paper.
We note that in CFT view, $c_I$ are related to the
anomalous dimensions $\gamma_I$ of chiral operators built from CFT-charged
superfields \cite{Nelson:2000sn}. At a superconformal fixed point,
$\gamma_I$ are determined simply by group theory factors, so are
quantized.

With appropriate values of the brane cosmological constants
at the orbifold fixed points ($y=0,\pi$), the ground state geometry
of the action
(\ref{5daction1}) is given by a slice of ${\rm AdS}_5$
having the following form of 5D metric:
\beq
\label{metric}
G_{MN}dx^Mdx^N=e^{-2kRy}\eta_{\mu\nu}dx^\mu dx^\nu+R^2dy^2.
\eeq
It is convenient to write the  5D action (\ref{5daction1}) in $N=1$
superspace\cite{Arkani-Hamed:2001tb,Marti:2001iw,Linch:2002wg}.
For the 5D SUGRA multiplet, we keep only the radion superfield
$$
T=\left(\,R+iB_5\,, \,\,\frac{1}{2}(1+\gamma_5)\Psi^{i=2}_5\,\right)\,
$$
and replace other fields by their vacuum expectation values. For
the 5D vector multiplets and hypermultiplets, one needs
appropriate $R$ and $B_5$-dependent field redefinitions to
construct the corresponding $N=1$ superfields
\cite{Marti:2001iw,Choi:2002wx}. After such field redefinition,
the relevant piece of the action is given by\footnote{
Note that our $H_I$ and $H^c_I$ differ from
\cite{Marti:2001iw} which is related to ours
by $H_I\rightarrow e^{-(\frac{3}{2}-{c}_I)Tk|y|}H_I$
and $H^c_I\rightarrow e^{-(\frac{3}{2}+{c}_I)Tk|y|}H^c_I$.}
\cite{Arkani-Hamed:2001tb,Marti:2001iw} \bea \label{5daction2}
&&
\int d^5x\,
\left[\,\int\,d^4\theta\,\frac{T+T^*}{2}
\left(\hat{H}^*_I\hat{H}_I+\hat{H}^{c*}_I\hat{H}^c_I\right)
+\left\{\int d^2\theta \frac{1}{4g_{5a}^2}TW^{a\alpha}W^a_{\alpha}
+h.c.\right\}\,\right]
\nonumber \\
=&&\int d^5x \,\left[\,\int d^4\theta
\,\frac{T+T^*}{2}\left(
e^{(\frac{1}{2}-{c}_I)(T+T^*)k|y|}H_I^*H_I
+e^{(\frac{1}{2}+{c}_I)(T+T^*)k|y|}H^{c*}_IH^{c}_I\right)
\right. \nonumber \\ && \quad\quad\quad\left. +\left\{\,
\int d^2\theta\, \frac{1}{4{g}_{5a}^2}TW^{a\alpha}W^a_{\alpha}
+h.c.\,\right\} \right]\,,\eea where $W^{a}_{\alpha}$ is the
chiral spinor superfield for the vector superfield ${\cal V}^a$
containing $(A^a_{\mu},\lambda^a)$ with
$\lambda^a=\frac{1}{2}(1-\gamma_5)\lambda^{a1}$,
${H}_I$ and ${H}_I^c$
are chiral superfields containing $(h_I^1,\psi_I)$ and
$(h_I^{2*},\psi^c_I)$, respectively, with
$\psi_I=\frac{1}{2}(1-\gamma_5)\Psi_I$,
$\bar{\psi^c}_I=\frac{1}{2}(1+\gamma_5)\Psi_I$.
Here we consider two superfields bases
for hypermultiplets, $(\hat{H}_I, \hat{H}^c_I)$
and $(H_I, H^c_I)$, which are related to each other
by
\bea
\hat{H}_I&\,=\,&e^{(\frac{1}{2}-c_I)Tk|y|}H_I,
\nonumber \\
\hat{H}^c_I&\,=\,&e^{(\frac{1}{2}+c_I)Tk|y|}H^c_I.
\nonumber
\eea

As the theory is
orbifolded by $Z_2: y\rightarrow -y$, all 5D fields should have a
definite boundary condition under $Z_2$.
The 5D SUGRA multiplet is assumed to have the standard
boundary condition
leaving the 4D $N=1$ SUSY unbroken.
To give a massless 4D gauge multiplet, the vector superfield 
${\cal V}^a$ is required to be $Z_2$-even:
\beq
{\cal V}^a(-y)={\cal V}^a(y).
\eeq
On the other hand, the hypermultiplet can have any $Z_2$-boundary
condition: 
\bea
H_I (-y) &\,=\,& z_I H_I (y),
\nonumber \\
H_I^c(-y) &\,=\,& -z_I H_I (y),
\eea
where $z_I=\pm 1$,
and then a massless
4D chiral superfield $Q_I$ originates from
either $H_I$ ($z_I=1$) or $H^c_I$ ($z_I=-1$).

In the superfield basis of $(H_I, H^c_I)$, 
the $I$-th 4D superfield
$Q_I$ appears as
the $y$-{\it independent} mode of 
$(H_I, H^c_I)$.
However in another superfield basis of
$(\hat{H}_I, \hat{H}^c_I)$ for which the 5D action
(\ref{5daction2}) has a standard form of K\"ahler
metric, the wavefunction of $Q_I$
appears to be {\it localized} near at one of the
orbifold fixed points:
\beq
\hat{H}_I \,\sim\, e^{-q_ITk|y|} \quad (z_I=1)
\quad
\mbox{or} 
\quad
\hat{H}^c_I \,\sim\, e^{-q_ITk|y|}\quad (z_I=-1),
\eeq
where
$$
q_I=z_Ic_I-\frac{1}{2}.
$$ 
So, $Q_I$ is localized near at $y=0$ if $q_I>0$,
while localized at $y=\pi$ if $q_I<0$.
As is well known, such localization of matter zero modes
can generate naturally the hierarchical Yukawa couplings.
It is in fact possible to achieve the dynamical localization
of matter zero modes in flat extra dimension \cite{Arkani-Hamed:1999dc,Arkani-Hamed:2001tb}. 
However in 5D SUGRA context,
the localization of matter zero modes
has the same origin as the localization
of gravity zero mode, i.e.
the gauging of graviphoton as (\ref{gauging}).
In this sense, warped extra dimension with
localized 4D gravity can be considered
as a more natural ground than flat extra dimension
for the dynamical localization of matter zero modes.
Furthermore, the localization of matter zero modes in
$\mbox{AdS}_5$  has an intrepretation
in terms of supersymmetric 4D CFT as the
renormalization group evolution at superconformal
fixed point \cite{Nelson:2000sn}.

In addition to the bulk action (\ref{5daction2}),
there can be brane actions at the fixed points
$y=0,\pi$.
The general covariance requires that
the 4D metric in brane action should be the
4D component of the 5D metric at fixed point.
Using the general covariance and also
the $R$ and $B_5$-dependent field redefinitions
which have been made to construct $N=1$ superfields,
one can easily find
the $T$-dependence of brane actions \cite{Marti:2001iw,Casas,Hebecker}.
For instance,
the brane actions which would be relevant for
Yukawa couplings and soft parameters
are given by\footnote{
The chiral anomaly of the $R$ and $B_5$-dependent field redefinition
induces $T$-dependent pieces in
$\omega_a$ and $\omega^\prime_a$ \cite{Choi:2002wx}.
But they are loop-suppressed and
not very relevant for the discussion in this paper.}
\bea
\label{braneaction}
S_{\rm brane}&=&\int\,d^5x\,
\left[\,\,\delta(y)\,\left\{
\int d^4\theta \,L_{I\bar{J}}(Z,Z^*)\Phi_I\Phi^*_J
\right.\right.\nonumber \\
&& \left.+\left(\,\int d^2\theta
\,\frac{1}{4} \omega_a(Z) W^{a\alpha}W^a_{\alpha}
+\lambda_{IJK}(Z)\Phi_I\Phi_J\Phi_K+{\rm h.c}\,\right)\,\right\}
\nonumber \\
&&+\,\,\delta(y-\pi)\left\{
\int d^4\theta \, e^{-(q_I\pi kT+q_J\pi kT^*)}
L^\prime_{I\bar{J}}(Z^\prime,Z^{\prime *})\Phi_I\Phi^{*}_J
\right.
\\
&&\left.\left.+\left(\,\int d^2\theta \,\frac{1}{4}
\omega_a^\prime(Z^\prime)
W^{a\alpha}W^a_{\alpha}
+e^{-(q_I+q_J+q_K)\pi kT}\lambda^\prime_{IJK}(Z^\prime)
\Phi_I\Phi_J\Phi_K
+{\rm h.c}\,\right)\,\right\}\,\right] \nonumber 
\eea
where $Z$ and $Z^\prime$ denote generic 4D
chiral superfields living {\it only} on the brane at $y=0$ and $y=\pi$,
respectively, and
$$\Phi_I=H_I \quad (z_I=1)
\quad
\mbox{or} 
\quad \Phi_I=H_I^c \quad (z_I=-1).
$$
Here $L_{I\bar{J}}$ ($L_{I\bar{J}}^\prime$) are generic hermitian functions
of $Z$ and $Z^*$ ($Z^\prime$ and $Z^{\prime *}$),
and $\omega_a$ and $\lambda_{IJK}$
($\omega^\prime$ and $\lambda^\prime_{IJK}$) are
generic holomorphic functions of $Z$ ($Z^\prime$).

If there is no light gauge-singlet 5D field
other than the minimal 5D SUGRA multiplet, 
SUSY breaking would be mediated by the radion superfield
$T$, and/or the brane superfields $Z,Z^\prime$, and/or
the 4D SUGRA multiplet which always participate in
the mediation of SUSY breaking through the conformal anomaly
\cite{anomaly}.
The relative importance of anomaly-mediation (compared
to radion-mediation) depends on the details of radion 
stabilization. Although anomaly-mediation
can give an important (or even dominant) contribution
to soft parameters in some special case \cite{Luty:2001jh}, 
in the class of models under consideration, the contributions
from anomaly-mediation are generically
negligible as they are further suppressed by warp factor as well as 
by the loop factor \cite{futurework}.
So in this paper, we will focus  on  
the soft SUSY breaking parameters
induced by the $F$-terms of
$T$ and  $Z,Z^\prime$.

The 4D Yukawa couplings and soft parameters can be
most easily studied by constructing the effective action
of massless 4D superfields.
As we have noted, in the superfield basis
of $({\cal V}^a, H_I, H^c_I)$, the {\it $y$-independent}
constant modes correspond to the massless 4D superfields
of the model.
Let $V^a$ denote the constant mode of
${\cal V}^a$, and
$Q_I$ to be the constant mode
of $H_I$ ($z_I=1$) or of $H_I^c$ ($z_I=
-1$). Here we will assume that all
visible 4D gauge and matter fields are in $\{V^a,Q_I\}$,
and examine their Yukawa couplings and soft SUSY breaking
parameters when the SUSY breaking is mediated
by $T$ and/or $Z,Z^\prime$.
Those Yukawa couplings and soft parameters
at the Kaluza-Klein scale
$M_{KK}\approx ke^{-\pi kR}$ can be evaluated from
the 4D effective action which can be written as
\bea
\label{4deffective1}
\left[\,\int d^4 \theta\,\, Y_{I\bar{J}}Q_IQ^*_J\,\right]
+\left[\,\int d^2 \theta
\,\left(\f{1}{4} f_a W^{a\alpha} W^a_\alpha + \tilde{y}_{IJK}Q_IQ_JQ_K\,\right)
 +\mbox{h.c.}\,
\right],
\eea
where $Y_{I\bar{J}}$ are hermitian wave function coefficients,
$f_a$ are holomorphic gauge kinetic functions,
and $\tilde{y}_{IJK}$ are holomorphic Yukawa couplings.
Using (\ref{5daction2}) and (\ref{braneaction}),
we find
\bea
\label{4deffective}
Y_{I\bar{J}}&=&\frac{1}{{q}_Ik}\left( 1- e^{-{q}_I\pi k(T+T^*)}
\right)\delta_{IJ}+L_{I\bar{J}}(Z,Z^*)
+\frac{L^\prime_{I\bar{J}}(Z^\prime,Z^{\prime *})
}{e^{({q}_I\pi kT+{q}_J\pi kT^*)}},
\nonumber \\
f_a&=&\frac{2\pi}{g_{5a}^2}T+\omega_a(Z)+\omega^\prime_a(Z^\prime),
\nonumber \\
\tilde{y}_{IJK}&=&\lambda_{IJK}(Z)+\frac{\lambda^\prime_{IJK}(Z^\prime)}{
e^{({q}_I+{q}_J+{q}_K)\pi kT}}. \eea 
Note that 5D SUSY in bulk
enforces that the Yukawa couplings of $Q_I$ originate entirely
from the brane action (\ref{braneaction}).

It is straightforward to compute soft parameters for
generic forms of $L_{I\bar{J}}, L^\prime_{I\bar{J}}$
\cite{futurework}.
Here we will consider the case with
\bea
L_{I\bar{J}}(Z,Z^*)&\,=\,&-\kappa_{I\bar{J}}ZZ^*,
\nonumber \\
L^\prime_{I\bar{J}}(Z^\prime,Z^{\prime *})&\,=\,&-\kappa^\prime_{I\bar{J}}
 Z^\prime Z^{\prime *},
\eea where $\kappa_{I\bar{J}}$ and $\kappa^\prime_{I\bar{J}}$ are generic
constants of order one.
However, we will keep $\omega_a$ and $\lambda_{IJK}$
to be generic holomorphic functions of $Z$, and 
$\omega^\prime_a$ and $\lambda^\prime_{IJK}$
to be generic holomorphic functions of $Z^\prime$.
The results for more general
forms of $L_{I\bar{J}}$ and $L^\prime_{I\bar{J}}$ will be presented elsewhere
\cite{futurework}. In regard to
the suppression of flavor-violating soft
parameters, those general results show a similar behavior as
our case. 
For simplicity, we further assume that 
$$\langle Z\rangle\ll 1,
\quad 
\langle Z^\prime \rangle \ll 1,
$$
in the unit with $M_5=1$, so
$$
\langle L_{I\bar{J}}\rangle \ll 1, \quad
\langle L^\prime_{I\bar{J}}
\rangle\ll 1.
$$
In regard to SUSY breaking,
we consider the general case in which 
any of the $F$-components of
$T,Z$ and $Z^\prime$ can be the major source of SUSY breaking. 
Note that here we are not concerned with
the dynamial origin of those $F$-components,
but with the resulting soft
parameters of visible fields for generic values of
the $F$-components.

Let $y_{IJK}$, $M_a$, $m^2_{I\bar{J}}$, and $A_{IJK}$
denote the Yukawa couplings, gaugino masses,
soft scalar masses, trilinear scalar couplings, 
respectively, 
for the {\it canonically normalized}
matter superfields $Q_I=\phi^I+\theta \psi^I+\theta^2 F^I$
and gauginos $\lambda^a$ which are renormalized at $M_{KK}$:
\bea
\frac{1}{2}y_{IJK}\phi_I\psi_J\psi_K-
\frac{1}{2}M_a\lambda^a\lambda^a-
\frac{1}{2}m^2_{I\bar{J}}\phi^I\phi^{J*}
-\frac{1}{6}A_{IJK}\phi^I\phi^J\phi^K+h.c.
\eea
One then finds from (\ref{4deffective}):
\bea
\label{yukawa}
y_{IJK} &&\,=\, (Y_IY_JY_K)^{-1/2}
\left(\lambda_{IJK}+\frac{\lambda^\prime_{IJK}}{e^{(q_I+q_J+q_K)\pi kT}}
\right),
\nonumber \\
M_a &&\,=\,\frac{F^T}{2R}+\frac{1}{2}g_a^2\left(
\frac{\partial \omega_a}{\partial Z}
F^Z+\frac{\partial \omega^\prime_a}{\partial Z^\prime}
F^{Z^\prime}\right)\,,
\nonumber \\
m^2_{I\bar{J}} &&\,=\, (Y_IY_J)^{-1/2}\left[
\,\frac{\pi^2 q_Ik\delta_{IJ}|F^T|^2}{e^{q_I\pi k(T+T^*)}-1}
+\kappa_{I\bar{J}}\left| F^Z \right|^2
+\frac{\kappa^\prime_{I\bar{J}}|F^{Z^\prime}|^2}{e^{(q_I\pi
kT+q_J\pi kT^*)}}\,\right],
\nonumber \\
A_{IJK} &&\,=\, -(Y_IY_JY_K)^{-1/2}\left[F^T
\frac{\partial}{\partial T}\ln
\left(\frac{\lambda_{IJK}+\lambda^\prime_{IJK}e^{-(q_I+q_J+q_K)\pi kT}}{
Y_IY_JY_K}\right)\right.
\nonumber \\
&&\left. \times \left(\lambda_{IJK}+\frac{\lambda^\prime_{IJK}}{
e^{(q_I+q_J+q_K)\pi kT}}\right)
+F^Z\frac{\partial \lambda_{IJK}}{\partial Z}
+\frac{F^{Z^\prime}}{e^{(q_I+q_J+q_K)\pi kT}}\frac{\partial
\lambda^\prime_{IJK}}{\partial Z^\prime}\,\right]\,,
\eea
where 
$$Y_I=\frac{1}{q_Ik}(1-e^{-q_I\pi k(T+T^*)}),
$$
$g_a^2$ are 4D gauge couplings,
and  $F^T,F^Z$ and $F^{Z^\prime}$ denote the $F$-components
of $T,Z$ and $Z^\prime$, respectively.

To be more concrete, let us consider the case that both the
Yukawa couplings and the brane SUSY breaking come entirely
from
$\lambda_{IJK}$ and $F^Z$ at $y=0$.
We further assume that 
the wavefunctions of 4D Higgs fields are localized
at $y=0$, which would be necessary to generate the top
quark Yukawa coupling of order one.
Let $Q_H$ and $Q_m$ denote the Higgs and quark/lepton superfields,
respectively, and specify only the quark/lepton flavor
indices ($m,n$) in the Yukawa couplings and trilinear $A$-parameters.
We then find
\bea
\label{result1}
y_{mn}&=&\frac{\lambda_{mn}}{\sqrt{Y_HY_mY_n}},
\nonumber \\
M_a&=&\frac{F^T}{2R}+\frac{1}{2}g_a^2
\frac{\partial \omega_a}{\partial Z}F^Z,
\nonumber \\
m^2_{m\bar{n}}&=&\frac{1}{\sqrt{Y_mY_n}}\left[\,
\frac{\pi^2 q_mk\delta_{mn}}{e^{2\pi q_mR}-1}\left|
{F^T}\right|^2+\kappa_{m\bar{n}}|F^Z|^2\,\right]
\nonumber \\
A_{mn}&=&-y_{mn}\left[
F^T\frac{\partial}{\partial T}\ln\left(
\frac{1}{Y_HY_mY_n}\right)+F^Z\frac{\partial}{
\partial Z}\ln(\lambda_{mn})\right]
\eea
where $\lambda_{mn}$ denote 
the holomorphic Yukawa couplings in the brane action (\ref{braneaction})
at $y=0$, and
\bea
\frac{1}{\sqrt{Y_H}}=
\left(\frac{1}{q_Hk}(1-e^{-q_H\pi k(T+T^*)})\right)^{-1/2}
\approx\,
&&
\sqrt{q_Hk} \quad~\quad~\quad~\quad~\quad~ (q_H>0)
\nonumber\\
\frac{1}{\sqrt{Y_m}}=
\left(\frac{1}{q_mk}(1-e^{-q_m\pi k(T+T^*)})\right)^{-1/2}
\approx \,
&& \cases{ 
\begin{array}{lc} 
\sqrt{q_mk} &\quad  (q_m>0) 
 \\  
~& \\
 \sqrt{|q_m|k}\,e^{-\pi |q_m|kR} &\quad  (q_m<0)
\end{array}}
\eea

As we have noted, $Q_m$ with $q_m<0$ is localized at $y=\pi$,
while $Q_m$ with $q_m>0$ is localized at $y=0$.
As a result, in the case that both the Yukawa couplings
and the brane SUSY breaking arise from  $y=0$,
the Yukawa couplings, $A$-parameters and squark/slepton masses 
(renormalized at $M_{KK}$)
involving the light quark/lepton superfields $Q_m$ with $q_m<0$  
are exponentially
suppressed as
\bea
\label{suppression}
y_{mn}&=&{\cal O}(e^{-(|q_m|+|q_n|)\pi kR})\,,
\nonumber \\
A_{mn}&=&{\cal O}(e^{-(|q_m|+|q_n|)\pi kR}M_a)\,,
\nonumber \\
m^2_{m\bar{n}}&=&{\cal O}(e^{-(|q_m|+|q_n|)\pi kR}M_a^2),
\eea
while those involving $Q_m$ with $q_m>0$ are unsuppressed.
On the other hand, there is no localization of $V^a$, so
no suppression of $M_a$.
Note that although the squark/slepton masses
from $F^Z$ are suppressed {\it only} for
$Q_m$ with $q_m<0$, 
the squark/slepton masses from $F^T$ are suppressed
for {\it all} $Q_m$ with $q_m\neq 0$.
This can be easily understood by noting that the radion
mediation is equivalent to the Scherk-Schwarz SUSY breaking
\cite{Kaplan:2001cg}.
The Scherk-Schwarz SUSY breaking 
corresponds to a twist of boundary condition, so its effects
are suppressed for any localized mode independently of
the position of localization.

In fact, for each model in which 
both the Yukawa couplings and the brane SUSY breaking
arise from $y=0$, there exists a dual model
in which the Yukawa couplings and the brane SUSY breaking
arise from $y=\pi$. 
Once we change the sign of all $q_I$ of the involved 
hypermultiplets, the dual model gives
the same 4D Yukawa couplings and soft parameters.
To see this,  let us consider the case that both the
Yukawa couplings and the brane SUSY breaking come entirely
from
$\lambda^\prime_{IJK}$ and $F^{Z^\prime}$ at $y=\pi$.
It is then straightforward to find
\bea
\label{result2}
y_{mn}&=&\frac{e^{-(q_H+q_m+q_n)\pi kR}\lambda^\prime_{mn}}{\sqrt{
Y_HY_mY_n}},
\nonumber \\
M_a&=&\frac{F^T}{2R}+\frac{1}{2}g_a^2\frac{\partial \omega^\prime_a}{
\partial Z^\prime}F^{Z^\prime},
\nonumber \\
m^2_{m\bar{n}}&=&\frac{1}{\sqrt{Y_mY_n}}\left[\,
\frac{\pi^2q_mk\delta_{mn}}{e^{2\pi q_mR}-1}|F^T|^2
+\kappa^\prime_{m\bar{n}}e^{-(q_m+q_n)\pi
  kR}|F^{Z^\prime}|^2\,\right], 
\nonumber \\
A_{mn}&=&-y_{mn}\left[\,F^T
\frac{\partial}{\partial T}
\ln\left(\frac{e^{-(q_H+q_m+q_n)\pi kT}}{Y_HY_mY_n}\right)+
F^{Z^\prime}\frac{\partial}{\partial Z^\prime}\ln(\lambda^\prime_{mn})
\,\right].
\eea
It is rather obvious that these results give
(\ref{result1}) when $(\lambda^\prime_{mn}, \kappa^\prime_{m\bar{n}},
Z^\prime, q_H, q_m)$
are replaced  by $(\lambda_{mn},\kappa_{m\bar{n}}, Z,
 -q_H, -q_m)$.
So in this case,
we need to localize 
the Higgs fields at $y=\pi$ by assuming
$q_H<0$ in order to have a large top quark Yukawa coupling,
and localize the light quark/lepton fields at $y=0$
by assuming $q_m>0$ in order to suppress the
Yukawa couplings and flavor-violating soft parameters.

The results of (\ref{result1}) and (\ref{result2}) show that
if the brane SUSY breaking takes place {\it only}
at one of the orbifold fixed points from which the Yukawa
couplings arise,
the flavor-violating soft parameters
of light squark/slepton generations
which are induced by $F^T$ and {\it one} of $F^Z, F^{Z^\prime}$
are suppressed as (\ref{suppression}).
So, the $\mbox{AdS}_5$ models with localized bulk matter
fields can avoid (or at least ameliorate) the SUSY flavor problem
while generating the hierarchical Yukawa couplings\footnote{
Another way to ameliorate the SUSY flavor problem
using extra dimension has been
suggested in \cite{Kubo:2002pv}.}.
Note that all of our results can be applied also
to models with {\it flat} extra dimension which
generate the hierarchical Yukawa couplings by localizing
the bulk matter fields.
For this, one can simply take the limit $k\rightarrow 0$
while keeping $q_Ik$ nonzero.
In the next section, we will consider the
radion-dominated SUSY breaking scenario in this framework,
which provides a concrete prediction on soft parameters,
and present some models which can pass the constraints on
flavor violation without severe fine tuning of parameters.

The results (\ref{yukawa}), (\ref{result1}) and
(\ref{result2}) are obtained 
from the 4D effective action (\ref{4deffective1}) without including
the possible threshold corrections due to massive
KK modes.
The (exponential) suppression
of $y_{mn}$ and $A_{mn}$ for light generations
is expected to be  stable against KK threshold corrections as 
it is due to the localization of zero modes in extra dimension.
In the 4D effective SUGRA point of view,
the suppression of $y_{mn}$ and $A_{mn}$ is a consequence of
an exponentially large K\"ahler metric $Y_{m\bar{n}}
\approx e^{2\pi |q_m|kR}\delta_{mn}/|q_m|k$
for the case of  (\ref{result1}),
or of an exponentially small holomorphic Yukawa couplings 
$\tilde{y}_{mn}=e^{-(|q_m|+|q_n|)\pi kR}\lambda^\prime_{mn}$ 
for the case of (\ref{result2}). 
These features also suggest that
the suppressions of $y_{mn}$ and $A_{mn}$ 
are stable againt KK threshold corrections.

On the other hand, $m^2_{m\bar{n}}$ generically get
threshold corrections of order $M_a^2/8\pi^2$.
However the geometric interpretaion
for the suppressed flavor-violating soft parameters
suggets that the flavor-violating part of the KK threshold correction
to $m^2_{m\bar{n}}$ is suppressed also by
$e^{-(|q_m|+|q_n|)\pi kR}$ as well as by the loop factor,
which means that the KK threshold correction gives 
just a subleading piece of flavor-violation in $m^2_{m\bar{n}}$.
There can be additional corrections to soft parameters
which are induced by non-renormalizable SUGRA interactions 
in 4D effective action \cite{choi}, 
but they are suppressed by
$M_{KK}^2/8\pi^2 M_{Pl}^2\sim e^{-2\pi kR}/8\pi^2$, so
are small enough.
There are also the model-independent SUSY breaking effects mediated
by the 4D superconformal  anomaly \cite{anomaly}.
In the models under consideration, the gravitino mass
is generically given by $m_{3/2}=
{\cal O}(e^{-\pi kR} M_a)$, so
the anomaly-mediated contributions to soft parameters are
$\delta M_a\sim e^{-\pi kR}M_a/8\pi^2$,
$\delta m^2_{m\bar{n}}\sim e^{-2\pi kR}M^2_a/(8\pi^2)^2$ 
and $\delta A_{mn}\sim e^{-\pi kR}M_a/8\pi^2$, which
are small enough.
So the leading radiative corrections to 
Yukawa couplings and soft parameters at $M_W$ come from
the standard renormalization group running down to $M_W$
starting from the boundary values at $M_{KK}$ given by
(\ref{result1}) or (\ref{result2}).

We note that the idea of AdS/CFT correspondence suggests 
a CFT framework which would reproduce the main
features  of our AdS models.
Indeed, models involving superconformal sector have been proposed
to generate hierarchical Yukawa couplings as well as
exponentially suppressed soft masses
\cite{Nelson:2000sn,Luty:2001jh,Karch:1998qa}.
It is then easy to see 
the correspondence: $q_I\pi kR
\rightarrow \gamma_I\ln (\Lambda/M_c)$, where
$\gamma_I$ is the anomalous dimension of $Q_I$
driven by the coupling to the SC sector,
and $\Lambda$ and $M_c$ are the cutoff scale and the decoupling scale
of the superconformal sector, respectively.
The $\mbox{AdS}_5$ models discussed here
provide a perturbative framework to generate the hierarchical
Yukawa couplings while suppressing the dangerous
flavor-violating soft parameters.

\section{Radion-dominated Scenario}

The results of (\ref{result1}) and (\ref{result2})
show that  potentially dangerous flavor-violating
soft parameters can be naturally suppressed in
$\mbox{AdS}_5$ models.
However the resulting soft parameters
still involve many adjustable free parameters,
particularly in the contributions from the brane SUSY breaking.
Obviously, the model becomes much more predictive
in the radion-dominated scenario\cite{Chacko:2000fn}
which we will discuss in somewhat detail
in this section.

To proceed, let us consider
the case that the Yukawa couplings come from
the brane action at $y=0$ and the Higgs zero modes are
localized at $y=0$.
To be definite, we assume that the
Kaluza-Klein scale is given by $M_{KK}=ke^{-\pi kR}
\sim 10^{16}$ GeV, however the results are not sensitive to
the precise value of $M_{KK}$.
In the radion-dominated scenario for this case, 
the Yukawa couplings and soft parameters
renormalized at $M_{KK}$ are
given by
\bea
\label{radiondomination}
y_{mn}&=&\frac{\lambda_{mn}\ln(1/\epsilon)}{\pi R\sqrt{Y_H}}
\sqrt{\frac{\phi_m\phi_n}{(1-\epsilon^{2\phi_m})(
1-\epsilon^{2\phi_n})}},
\nonumber \\
M_a&=& \frac{F^T}{2R},\nonumber \\
A_{mn}&=&2y_{mn}\ln(1/\epsilon)\left(
\frac{\phi_m}{\epsilon^{-2\phi_m}-1}+
\frac{\phi_n}{\epsilon^{-2\phi_n}-1}
+\frac{\phi_H}{\epsilon^{-2\phi_H}-1}\right)\frac{F^T}{2R},
\nonumber \\
m^2_{m\bar{n}}&=&\delta_{mn}\left(\,
2\ln(1/\epsilon)\frac{\phi_m}{\epsilon^{\phi_m}-
\epsilon^{-\phi_m}}\left|\frac{F^T}{2R}\right|\,\right)^2,
\eea
where
$\epsilon\approx 0.2$ denotes the Cabbibo angle,
$Y_H\approx 1/\sqrt{q_Hk}$,
 and
\bea
\phi_m&=&\frac{q_m\pi kR}{
\ln(1/\epsilon)}=\left(z_mc_m-\frac{1}{2}\right)
\frac{\pi kR}{
\ln(1/\epsilon)},
\nonumber \\
\phi_H&=& \frac{q_H\pi kR}{\ln(1/\epsilon)}=
\left(z_Hc_H-\frac{1}{2}\right)
\frac{\pi kR}{
\ln(1/\epsilon)},
\nonumber 
\eea
for the kink mass $c_Ik$   and the $Z_2$-boundary condition
factor $z_I(=\pm 1)$ for the 5D hypermultiplets ($H_I, H^c_I$)
whose zero modes correspond to the quark/lepton superfields
($I=m$)
or the Higgs superfields ($I=H$). 
As we have noted, $c_I$ and thus $\phi_I$ are likely
to be quantized. The observed quark and lepton masses
indicate that $\phi_m$ are integers for $\epsilon\approx 0.2$,
so that all Yukawa couplings are given by
an integer power of $\epsilon$ in their order of magnitudes.

Let $\psi_m=\{q_i, u_i, d_i, \ell_i, e_i\}$ 
($i=1,2,3$) denote the
known three generations of
the left-handed quark-doublets ($q_i$), up-type 
antiquark-singlets ($u_i$), down-type antiquark singlets
($d_i$), lepton-doublets ($\ell_i$), and anti-lepton singlets 
($e_i$).  The Yukawa 
couplings can be written as
\bea
{\cal L}_{\rm Yukawa}=
 y^u_{ij}H_2q_iu_j+y^d_{ij}H_1q_id_j+y^\ell_{ij}H_1\ell_ie_j
\eea
and the squark/sleptons $\phi_m=\{\tilde{q}_i,\tilde{u}_i, \tilde{d}_i,
\tilde{\ell}_i, \tilde{e}_i\}$ 
have the soft SUSY breaking couplings:
\bea
{\cal L}_{\rm soft}&=&-\left(\,
 A^u_{ij}H_2\tilde{q}_i\tilde{u}_j+A^d_{ij}H_1\tilde{q}_i\tilde{d}_j
+A^\ell_{ij}H_1\tilde{\ell}_i\tilde{e}_j
\right.\nonumber \\
&&\left.+m^{2(\tilde{q})}_{i\bar{j}}\tilde{q}_i
\tilde{q}^*_j
+m^{2(\tilde{u})}_{i\bar{j}}\tilde{u}_i
\tilde{u}_j^*
+m^{2(\tilde{d})}_{i\bar{j}}\tilde{d}_i
\tilde{d}_j^*
+m^{2(\tilde{\ell})}_{i\bar{j}}\tilde{\ell}_i
\tilde{\ell}^*_j
+m^{2(\tilde{e})}_{i\bar{j}}\tilde{e}_i
\tilde{e}_j^*\,\right)
\eea

Let us first consider the quark/squark sector.
There can be several different choices of $\phi_m\equiv
\phi(Q_m)$ which would yield the observed quark masses and mixing
angles \cite{Chun:1996xv}. Here we will consider one example:
\bea
&& \phi(H_1)=\phi(H_2)=2,
\nonumber \\
&&
\phi(q_i)=(-3,-2,2),\nonumber \\
&& \phi(u_i)=(-5,-2,2),
\nonumber \\
&&
\phi(d_i)=(-3,-2,-2),
\eea
for which $\tan\beta\sim \epsilon^2 m_t/m_b$ is not large.
These values of $\phi(Q_m)$  give the following forms of
Yukawa coupling matrices
\bea
y^u_{ij}&\,=\,& \pmatrix{ \epsilon^8\lambda^u_{11} & \epsilon^5
\lambda^u_{12} & \epsilon^3\lambda^u_{13} \cr
\epsilon^7\lambda^u_{21} & \epsilon^4\lambda^u_{22} & 
\epsilon^2\lambda^u_{23} \cr
\epsilon^5\lambda^u_{31} & \epsilon^2\lambda^u_{32} & 
\lambda^u_{33}}\,,
\nonumber \\
y^d_{ij}&\,=\,& \pmatrix{\epsilon^6\lambda^d_{11} & \epsilon^5\lambda^d_{12}
 &\epsilon^5\lambda^d_{13} \cr
\epsilon^5\lambda^d_{21} & \epsilon^4\lambda^d_{22} & \epsilon^4\lambda^d_{23}
 \cr
\epsilon^3\lambda^d_{31} & \epsilon^2\lambda^d_{32} &
\epsilon^2\lambda^d_{33}}\,,
\eea
where $\lambda^u_{ij}$ and $\lambda^d_{ij}$ are the
coefficients of order unity\footnote{Here $\lambda^u_{ij}$
and $\lambda^d_{ij}$ are redefined from
$\lambda_{mn}$ in (\ref{radiondomination}) to 
include the coefficients of order one depending
on $\ln(1/\epsilon)/\pi R$,$\phi(q_i)$,
$\phi(u_i)$, and $\phi(d_i)$.}. 
The soft parameters renormalized at $M_{KK}$ are determined to 
be\footnote{Here we have ignored small
corrections suppressed by high powers of $\epsilon$.}
\bea
\frac{A^u_{ij}}{y^u_{ij}}\,&=&\,
2M_{1/2}\ln 5\pmatrix{ 8 & 5 & 3\cr
7 & 4 & 2\cr
5 & 2 & {\cal O}(\epsilon^4)}\,,
\nonumber \\
\frac{A^d_{ij}}{y^d_{ij}}\,&=&\,
2M_{1/2}\ln 5\pmatrix{ 6& 5 & 5\cr
5& 4 & 4\cr
3& 2 & 2}\,,
\nonumber 
\eea
\bea
m^{2(\tilde{q})}_{i\bar{j}}&&\,=\,
(2\ln 5)^2|M_{1/2}|^2\,\pmatrix{
9\epsilon^6 &0&0\cr
0& 4\epsilon^4&0\cr
0&0& 4\epsilon^4}\nonumber \\
&&\,\approx\,
|M_{1/2}|^2\,\pmatrix{
6\times 10^{-3}&0&0\cr
0& 6\times 10^{-2}&0\cr
0&0& 6\times 10^{-2}}
\nonumber \\
m^{2(\tilde{u})}_{i\bar{j}}&&\,=\,
(2\ln 5)^2|M_{1/2}|^2\,\pmatrix{25\epsilon^{10}&0&0\cr
0& 4\epsilon^4&0\cr
0&0& 4\epsilon^4}\nonumber \\
&&\,\approx\, |M_{1/2}|^2\,\pmatrix{
3\times 10^{-5}&0&0\cr
0& 6\times 10^{-2}&0\cr
0&0& 6\times 10^{-2}}\,,
\nonumber \\
m^{2(\tilde{d})}_{i\bar{j}}&&\,=\,
(2\ln 5)^2|M_{1/2}|^2\,\pmatrix{9\epsilon^6&0&\cr
0&4\epsilon^4&0\cr
0&0&4\epsilon^4}\nonumber \\
&&\,\approx\,
|M_{1/2}|^2\,\pmatrix{6\times 10^{-3}&0&0\cr
0& 6\times 10^{-2}&0\cr
0&0&6\times 10^{-2}}\,,
\eea
where $M_{1/2}=F^T/2R$ denotes the universal gaugino
mass at $M_{KK}$.
The above Yukawa coupling matrices  produce
well the observed quark masses and mixing angles.
Also, after taking into account the renormalization group evolution
from $M_{KK}$ to the weak scale $M_W$,
the resulting soft parameters  pass 
all phenomenological constraints (including those
from flavor-changing processes)
for a reasonable range of $M_{1/2}$, e.g. $M_{1/2}\gtrsim 300$ GeV,
and for generic values of $\lambda^{u}_{ij}$ and $\lambda^d_{ij}$ 
which are taken to be of order unity \cite{futurework}.

One distinctive feature of the radion-mediated SUSY breaking
(for dynamically localized matter superfields) is 
that it gives a rather large
value of {\it non-universal} $A_{mn}/y_{mn}$
which is of the order of $M_{1/2}\ln y_{mn}$.
As a consequence, the most stringent constraint on
the model comes from $\mu\rightarrow e\gamma$ which would
be induced by the slepton $A$-coupling: $A^\ell_{ij}H_1\tilde{\ell}_i
\tilde{e}_j$.
To satisfy the experimental bound on $\mbox{Br}(\mu\rightarrow
e\gamma)$ for generic models of radion-mediation, 
one would need some degree of fine tuning 
for some of the off-diagonal elements 
of the lepton Yukawa matrix {\it unless}
the slepton and gaugino masses are heavier than {\cal O}(1) TeV.
Another distinctive feature of radion-mediation
is that although $A_{mn}/y_{mn}$ are highly flavor-dependent,
they are quantized up to small corrections being
a high power of $\epsilon$ (see Eq. (\ref{radiondomination}).
This feature provides a natural way to 
minimizes the  $\mu\rightarrow e\gamma$ rate, 
which is to choose
\beq
\phi(\ell_1)=\phi(\ell_2)\quad
\mbox{or}
\quad
\phi(e_1)=\phi(e_2).
\eeq

Let us thus consider the following 
two examples for the lepton/slepton sector:
\bea
\mbox{Model I}: \quad
&& \phi(\ell_i)=(-3,-3,-1),
\nonumber \\
&& \phi(e_i)=(-4,-1,-1),
\nonumber \\
\mbox{Model II}:\quad
&& \phi(\ell_i)=(-5,-2,-1),
\nonumber \\
&& \phi(e_i)=(-2,-2,-1).
\eea
The model I gives at $M_{KK}$:
\bea
\label{model1}
y^{\ell}_{ij}&=&\pmatrix{\epsilon^7\lambda^\ell_{11}&
\epsilon^4\lambda^\ell_{12}&\epsilon^4\lambda^\ell_{13}\cr
\epsilon^7\lambda^\ell_{21}&\epsilon^4\lambda^\ell_{22}&
\epsilon^4\lambda^\ell_{23}\cr
\epsilon^5\lambda^\ell_{31}&\epsilon^2\lambda^\ell_{32}
&\epsilon^2\lambda^\ell_{33}},
\nonumber \\
\frac{A^\ell_{ij}}{y^\ell_{ij}}&=&
2M_{1/2}\ln 5\pmatrix{7&4&4\cr
7&4&4\cr
5&2&2},
\nonumber \\
m^{2(\tilde{\ell})}_{i\bar{j}}&=&
(2\ln 5)^2|M_{1/2}|^2\pmatrix{9\epsilon^6&0&0\cr
0&9\epsilon^6&0\cr
0&0&\epsilon^2}
\nonumber \\
&\approx& \left| M_{1/2} \right|^2 \,\pmatrix{ 6\times 10^{-3} &0  & 0 \cr
                    0 & 6\times 10^{-3} & 0 \cr 
                    0 & 0 & 0.4 }
\nonumber \\
m^{2(\tilde{e})}_{i\bar{j}}&=&
(2\ln 5)^2|M_{1/2}|^2\pmatrix{
16\epsilon^8&0&0\cr
0&\epsilon^2&0\cr
0&0&\epsilon^2}
\nonumber \\
&\approx& 
\left|M_{1/2}\right|^2 \,\pmatrix{ 4 \times 10^{-4} & 0 & 0 \cr
          0 & 0.4 & 0 \cr
          0 & 0 & 0.4 }
\eea
while the model II gives
\bea
\label{model2}
y^{\ell}_{ij}&=&\pmatrix{
\epsilon^7\lambda^\ell_{11}&\epsilon^7\lambda^\ell_{12}&
\epsilon^6\lambda^\ell_{13}\cr
\epsilon^4\lambda^\ell_{21}&\epsilon^4\lambda^\ell_{22}&
\epsilon^3\lambda^\ell_{23}\cr
\epsilon^3\lambda^\ell_{31}&\epsilon^3\lambda^\ell_{32}&
\epsilon^2\lambda^\ell_{33}},
\nonumber \\
\frac{A^\ell_{ij}}{y^\ell_{ij}}&=&
2M_{1/2}\ln 5\pmatrix{7&7&6\cr
4&4&3\cr
3&3&2},
\nonumber \\
m^{2(\tilde{\ell})}_{i\bar{j}}&=&
(2\ln 5)^2|M_{1/2}|^2\pmatrix{25\epsilon^{10} &0&0\cr
0&4\epsilon^4&0\cr
0&0&\epsilon^2}
\nonumber \\
&\approx& \left| M_{1/2} \right|^2 \, 
\pmatrix{ 
  3 \times 10^{-5} & 0 & 0 \cr
  0 & 6 \times 10^{-2} & 0 \cr 
  0 & 0 & 0.4 },
\nonumber \\
m^{2(\tilde{e})}_{i\bar{j}}&=&
(2\ln 5)^2|M_{1/2}|^2\pmatrix{4\epsilon^4&0&0\cr
0&4\epsilon^4&0\cr
0&0&\epsilon^2}
\nonumber \\
&\approx& \left| M_{1/2} \right|^2 \, 
\pmatrix{
  6 \times 10^{-2} & 0 &0 \cr
  0 & 6\times 10^{-2} & 0 \cr 
  0 & 0 & 0.4 }.
\eea

The above results pass all phenomenological constraints 
{\it except for} $\mu\rightarrow e\gamma$
when all $\lambda^\ell_{ij}$ are taken to be of order
unity. To make the models to be consistent with
$\mu\rightarrow e\gamma$, one needs some degree of fine tuning
for $\lambda^\ell_{12}$ and $\lambda^\ell_{21}$
defined in (\ref{model1}) and (\ref{model2}).
Through a detailed numerical analysis including the renormalization
group evolution of soft parameters from $M_{KK}\sim 10^{16}$
GeV to $M_W$, we find that $\lambda^\ell_{12}$ and $\lambda^\ell_{21}$
are required to satisfy
\bea
\mbox{Model I}:\quad
&& \lambda^\ell_{12}\lesssim 2\times 10^{-2}\left(
\frac{M_{1/2}}{500\,\, \mbox{GeV}}\right)^2,
\nonumber \\
&& \lambda^\ell_{21}\lesssim 10^{-1} \left(
\frac{M_{1/2}}{500\,\, \mbox{GeV}}\right)^2,
\nonumber \\
\mbox{Model II}: \quad
&&\lambda^\ell_{12} \lesssim 5\times 10^{-2}\left(\frac{
M_{1/2}}{500 \,\,\mbox{GeV}}\right)^2,
\nonumber \\
&&\lambda^\ell_{21}\lesssim 10^{-2}\left(
\frac{M_{1/2}}{500\,\,\mbox{GeV}}\right)^2.
\eea
This can be considered to be a fine-tuning of parameters,
but not a serious one.

\section{Conclusion}

In this paper, we have examined the soft SUSY
breaking parameters in supersymmetric theories on a slice of
$\mbox{AdS}_5$ 
in which the hierarchical Yukawa couplings
are generated by  localizing the bulk matter fields
in extra dimension.
In this class of models, the localization of matter
zero modes has the common origin as the localization
of gravity zero mode, i.e. the gauging of graviphoton
which gives nonzero bulk cosmological constant
and hypermultiplet masses.
As a result, generically the small Yukawa couplings are given
by $y\sim e^{-c\pi kR}$ for a constant $c$ of order unity,
and so the radion $R$ is required to be stabilized at 
$e^{-\pi kR}\approx 10^{-2}-10^{-5}$.
Unless there is an additional gauge-singlet 5D field
other then the minimal 5D SUGRA multiplet, SUSY
breaking  is mediated by
the radion superfield in bulk, and/or some brane superfields
at $y=0$ and $\pi$, and/or the 4D SUGRA multiplet in bulk which
participates always in SUSY breaking through
the conformal anomaly.
Although it depends on the details of the radion stabilization,
generically the anomaly-mediated gaugino mass
in the models under consideration
is suppressed by a small warp factor (compared to
to the radion-mediated gaugino mass)
as well as by the small loop factor, so can be ignored.
We then found that if the brane SUSY breaking takes
place {\it only} at one of the orbifold fixed points
from which the Yukawa couplings arise, which is a natural
setting, 
the dangerous flavor-violating soft parameters
can be naturally suppressed by some powers of
the warp factor, so the model can avoid (or ameliorate)
the SUSY flavor problem.
This type of models can be considered as the AdS dual of
the recently studied 4D models containing a 
supersymmetric CFT to generate hierarchical Yukawa couplings.
If SUSY breaking is dominated by the radion-mediation,
the model provides a concrete prediction on soft
parameters which can be tested by low energy
experiments. We have presented some models in this framework
which pass all phenomenological constraints from
flavor violating processes without severe fine tuning
of parameters.

\bigskip
\noindent
{\bf Acknowledgments}

KC, DYK and IWK are supported by KRF PBRG 2002-070-C00022
and TK is supported by the Grants-in-Aid for the Promotion
of Science No. 14540256 from the Japan Society
for the Promotion of Science.

\end{document}